# Diving deep into the Milky Way using Anti-Reflection Coatings for Astronomical CCDs

Anmol Aggarwal[a†], Ashi Mittal[a†], George M. Seabroke[b], Nitin K. Puri[a]*

[a] Advanced Sensor Laboratory & Nanomaterials Research Laboratory, Department of Applied Physics, Delhi Technological University, Delhi, 110042, India.

[b] Mullard Space Science Laboratory, Department of Space & Climate Physics, Faculty of Maths & Physical Sciences, University College, London, RH5 6NT, United Kingdom.

[†] Authors have equal contributions to this research

[*] Corresponding author, Email IDs: nitinkumarpuri@dtu.ac.in, nitinpuri2002@yahoo.co.in

## Abstract

We report two anti-reflection (AR) coatings that give better quantum efficiency (QE) than the existing AR coating on the Gaia astrometric field (AF) CCDs. Light being the core of optical astronomy is extremely important for such missions, therefore, the QE of the devices that are used to capture it should be substantially high. To reduce the losses due to the reflection of light from the surface of the CCDs, AR coatings can be applied. Currently, the main component of the Gaia satellite, the AF CCDs use hafnium dioxide ($HfO_2$) AR coating. In this paper, the ATLAS module of the SILVACO software has been employed for simulating and studying the AF CCD pixel structure and several AR coatings. Our findings evidently suggest that zirconium dioxide ($ZrO_2$) and tantalum pentoxide ($Ta_2O_5$) will prove to be better AR coatings for broadband astronomical CCDs in the future and will open new avenues for understanding the evolution of the Milky Way.

## 1. Introduction

To enrich astronomy with more precise measurements, the European Space Agency (ESA) launched the Gaia Satellite in the L2 orbit in December 2013. Its two telescopes are keeping an eye on millions of stars, galaxies, and solar system objects to produce high-precision astrometric and spectroscopic measurements (Prusti et al. 2016). The continuous scan gives data sets that are repeatedly reduced to calculate the parallax, position, and proper motion of the celestial objects that are observed by the satellite (Crowley et al. 2016). The focal plane of the Gaia satellite contains 106 custom-built charged coupled devices (CCDs). These CCDs were designed and manufactured by e2v technologies, United Kingdom (Walker et al. 2008).

Gaia CCDs were fabricated by e2v in three different variants –astrometric field (AF), red photometer (RP) and blue photometer (BP); each of these are optimised for different wavelength ranges (De Bruijne 2012). The AF CCDs are built using silicon (Si) as a substrate with an anti-reflection (AR) coating which has a maximum photon absorption for a light of 650 nm; this CCD has an extensive wavelength detection range of 330-1050 nm (Seabroke et al. 2008). There are 78 AF CCDs on the Gaia Focal plane which are 16 μm thick (Crowley et al. 2016). The BP and RP are enhanced CCDs that have exceptional sensitivity towards the blue (330-680 nm) and red (640-1050 nm) regions of the light spectrum respectively. 7 BP CCDs present on the Gaia focal plane, have the maximum photon absorption for the light of 360 nm because of its AR coating (Crowley et al. 2016). Correspondingly, 7 RP CCDs have the maximum photon absorption for light of 750 nm (Crowley et al. 2016). These CCDs have an image area of 4500 lines × 1966 columns (here, lines and columns, refer to the rows and columns of the pixels of the CCD respectively) (De Bruijne 2012). A schematic diagram of the arrangement of CCDs on the actual Gaia focal plane is shown in **Fig. 1**. The detectors are operating in the time delay and integration mode with a period of 982.8 μs, which synchronises their line transfer rate with the satellite rotation rate (Crowley et al. 2016; De Bruijne 2012). In Gaia parlance, line transfer means electrons being transferred in a row of pixels (Crowley et al. 2016).

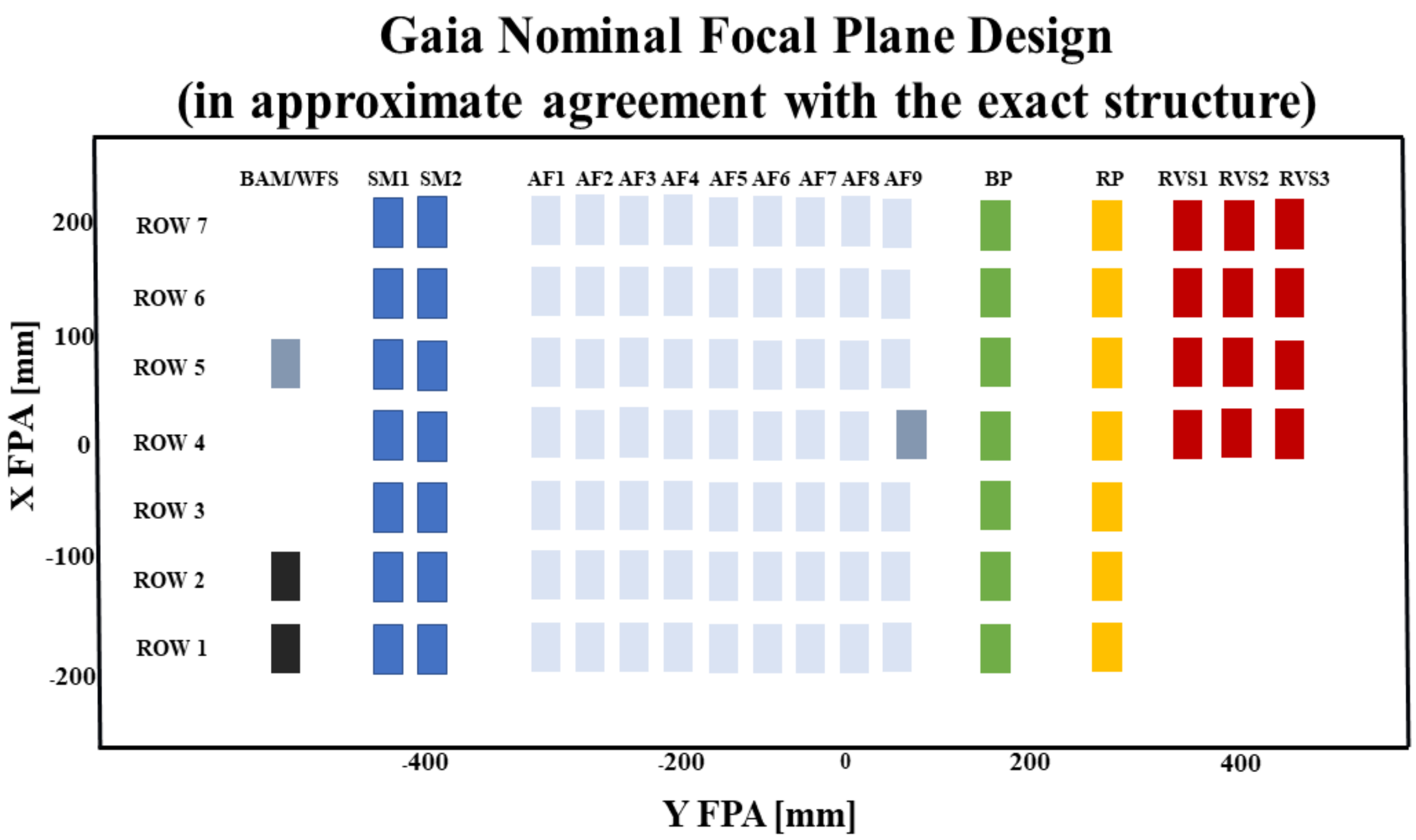

**Fig. 1:** The Gaia Focal Plane has 106 CCDs arranged in seven different rows. The red rectangles represent radial velocity spectrometers (RVS). The green and yellow rectangles depict blue and red photometers (BP and RP) respectively, while the grey and black rectangles show wavefront sensors (WFS) and basic angle monitors (BAM) respectively. The light blue rectangles depict the most abundant i.e., AF CCDs and the dark blue rectangles are called sky mappers (SM). The SM and the WFS are also AF CCDs. The RVS and BAM have a construction similar to the RP.

As the name suggests, AR coatings are essentially important in improving the internal quantum efficiency (IQE) of the CCDs because they play a vital role in enhancing their photon absorption by reducing the losses due to the reflection of the incident light. The IQE is used as a measuring stick to determine the performance of an AR coating which is the ratio of the number of photons that enter the CCD to the number of photons incident on the device (Devi et al. 2012). The external quantum efficiency (EQE) is the ratio of the number of photoelectrons generated to the number of photons incident on the CCD (Devi et al. 2012). Since the Gaia CCDs generally do not get a lot of time to capture the incoming light from distant objects, their IQE should be substantially high. An AR coating is a thin layer of high refractive index material (Lesser 1987). It also acts as a line of defence for the CCDs (Lesser 1994).

The Gaia AF CCDs have been analysed since they cover the maximum area (78 out of 106 CCDs) of the Gaia focal plane and capture the widest wavelength range. Studies that amalgamate AR coatings and astronomical CCDs have been conducted. The IQE of the device has been deliberated, which in fact is directly proportional to the EQE of the device. Several AR coatings for the Gaia AF CCDs have been scrutinized to enhance this factor using the SILVACO software.

## 2. Simulations and Calculations

SILVACO TCAD is used to simulate several electronic and optical devices. It uses numerical methods for simulations so that the development and optimisation of such devices can be expedited. For simulating the effects of AR coatings on an optoelectronic device, the LUMINOUS and the ATLAS modules of this software can be used.

Since the Gaia AF CCDs are back-illuminated devices, the AR coatings are applied on the backside. Photons are also fired on this side in our simulations. The operating voltage was set to 10V as suggested by Seabroke et al. (2009). In SILVACO simulations the ratio of available current density ($J_{available}$) and source current density ($J_{source}$) gives the IQE of a device. The number of photoelectrons generated in a device can be determined using the electron current density ($J_n$).

We simulated an AF CCD pixel in 2-dimensions (2D) using SILVACO, our aim is to obtain the IQE of the device with various AR coatings at different wavelengths. We assessed several materials as AR coatings namely, hafnium dioxide ($HfO_2$), aluminium oxide ($Al_2O_3$), zinc sulphide (ZnS), zirconium dioxide ($ZrO_2$), and tantalum pentoxide ($Ta_2O_5$). Although an AR coating of $HfO_2$ is already present on the AF CCDs (Short et al. 2005), which gives impressive results, we suspected that there was a scope of improvement in the IQE of the devices. Our simulations proclaim that the EQE of the Gaia AF CCDs with this AR coating peaks at 650 nm with a value of 0.99. Simulating these CCDs without any AR coating revealed that there is a large increase of 37.2% in the IQE when AR coatings are applied on such devices. Hence, AR coatings can be regarded as a cornerstone of astronomical CCDs.

A CCD is essentially a MOS-based device which acts as an image detector. In MOS, the letter ‘M’ stands for metal, the letter ‘O’ stands for oxide and the letter ‘S’ stands for semiconductor (Lundström 1981). Therefore, the structure of the CCD consists of a layer of metallic electrodes placed on a thin insulating layer lined over a semiconductor substrate. Lundström (1981) reports the details of the MOS structure. CCDs are of two types, front-illuminated and back-illuminated. In the former, light is incident on the electrodes, while the light is incident on the semiconductor substrate in the latter (Lesser 2014, 2015). The back-illuminated CCDs are constructed in actual practice because they offer a larger surface area for photogeneration and photon absorption than front-illuminated CCDs (Lesser 2015).

For simulating the Gaia AF CCD pixel in SILVACO several structural parameters were required which were derived from (Seabroke, Holland et al. 2008, 2009, 2010; Seabroke, Prod'homme et al. 2010). The pixel structure of the CCD has three different faces, which is evident from **Fig. 2**. We simulated the structure in the across scan (AC) direction in 2D with a pixel size of 16 μm × 30 μm. The 16 μm thickness of the pixel is subdivided into multiple layers, the details of which are summarised in **Table 1**.

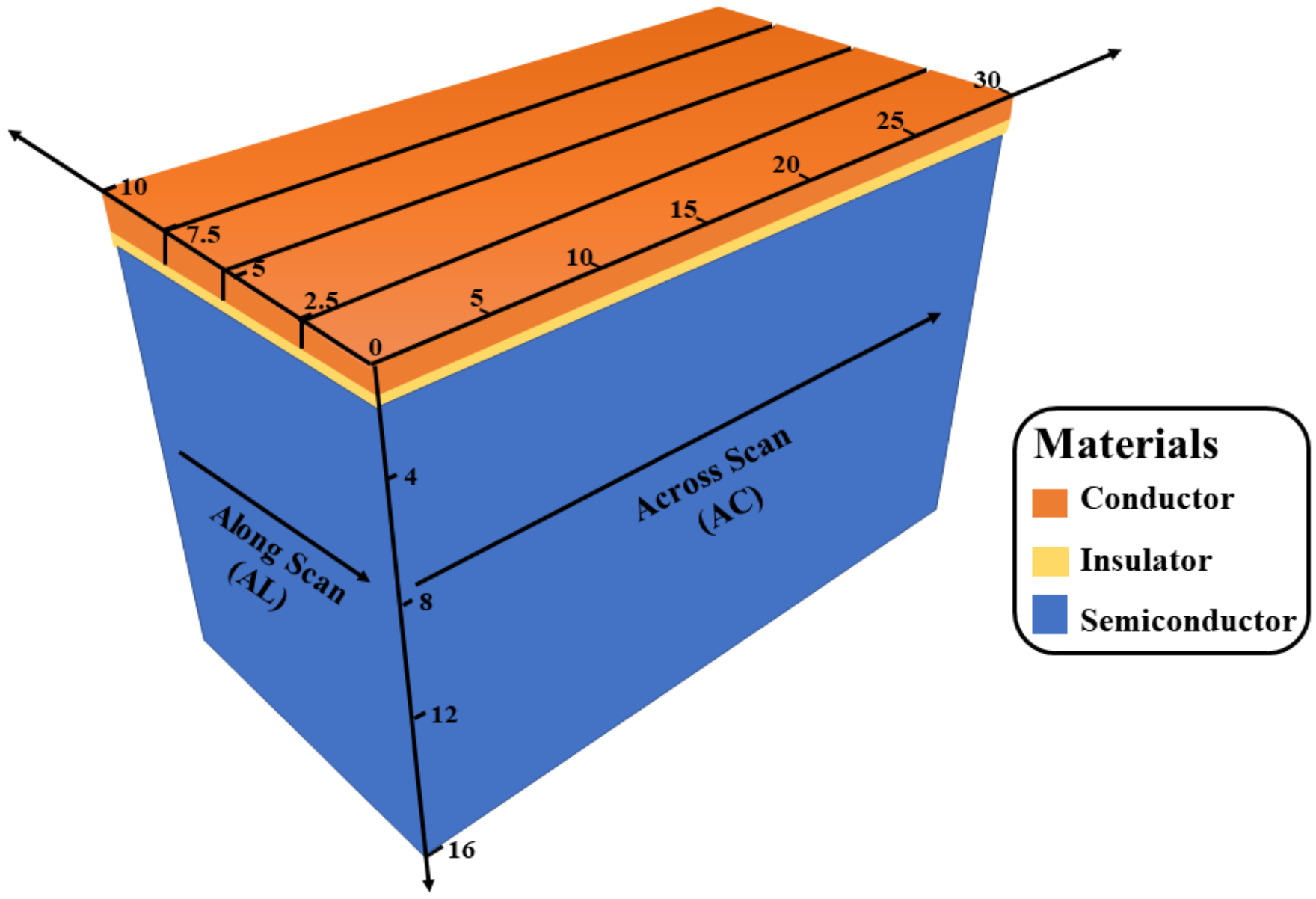

**Fig. 2:** Structure of a single pixel of the Gaia AF CCD (all the measurements are in μm).

**Table 1:** The thickness and doping densities of different layers in the AF CCD pixel structure (Seabroke et al. 2009).

| S. No. | Material | Type | Thickness (μm) | Doping Density ($cm^{-3}$) |
|---|---|---|---|---|
| **1** | Polycrystalline Si | Conductor | 0-0.5 | 0 |
| **2** | Silicon Dioxide ($SiO_2$) | Insulator | 0.5-0.63 | 0 |
| **3** | N-type Si | Semiconductor | 0.63- 1.17 | $2.65 \times 10^{16}$ |
| **4** | P-type Si | Semiconductor | 1.17-16 | $1.3 \times 10^{14}$ |

The Gaia AF CCD entails a lot of interesting structural intricacies, namely, buried channel (BC), supplementary buried channel (SBC) and anti-blooming drain (ABD) (Seabroke et al. 2008). A BC is formed near a p-n junction by placing a thin layer of negatively doped Si on a positively doped substrate. This channel has no free charges and is used to hold the photogenerated electrons before they are read out. A photon incident on a CCD creates an electron-hole pair; the electron so generated, then moves to the BC.

The SBC is a support feature to the BC but is no less functional. When the number of photo-generated electrons exceeds the capacity of the SBC they spill into the BC which holds them until they are read out (Seabroke et al. 2010, 2013). The length of the BC and the SBC runs from 4.5 to 29 µm in the across scan (AC) direction (Seabroke, Prod'homme et al. 2010). The ABD, which is present on either side of the pixel in the AC direction is a shielding feature. It prevents the electrons from divulging into the adjacent pixel (Seabroke et al. 2010). Now to simulate all these interesting features in SILVACO, the whole AF CCD image pixel was modelled with uniform doping as suggested by (Seabroke et al. 2009).

The Gaia AF CCDs work at a temperature of 163 K (Seabroke et al. 2008) to minimize the dark current, which is attributed to the false positive signals produced in the CCD due to the thermal energy of the electrons (Lesser 2015). Therefore, to match our theoretical simulations with the experimental results reported by (Walker et al. 2008), we conducted our simulations at the same temperature.

In order to simulate the CCD pixel structure, the SILVACO ATLAS package uses some constants for the simulated materials that are handled internally. It uses the Monte Carlo method to model the structure. **Table 2** compares the experimental values of the constants with the values used during our simulations.

**Table 2:** List of theoretical and experimental constants.

| S. No. | Name of the Constant | Theoretical Values used in SILVACO ATLAS | Experimental Values |
|---|---|---|---|
| 1. | Energy gap of Si ($E_g$) | 1.08 eV at 300 K<br>1.11 eV at 163 K | 1.12 eV at 300 K (Strehlow and Cook 1973) |
| 2. | Relative Permittivity of Si ($\varepsilon_r$) | 11.8 at 300 K<br>11.8 at 163 K | 11.66 at 300 K (Krupka, Breeze, Centeno et al. 2006) |
| 3. | Electron Affinity of Si (X) | 4.17 eV at 300 K<br>4.16 eV at 163 K | 4.05 eV at 300 K (Melnikov and Chelikowsky 2004; Chanana 2022) |
| 4. | Temperature (T) | 163 K | 163 K (Seabroke et al. 2008) |

**Fig. 3** exhibits the simulated structure of the Gaia AF CCD pixel; the illustration was generated using the Tonyplot tool of the SILVACO software.

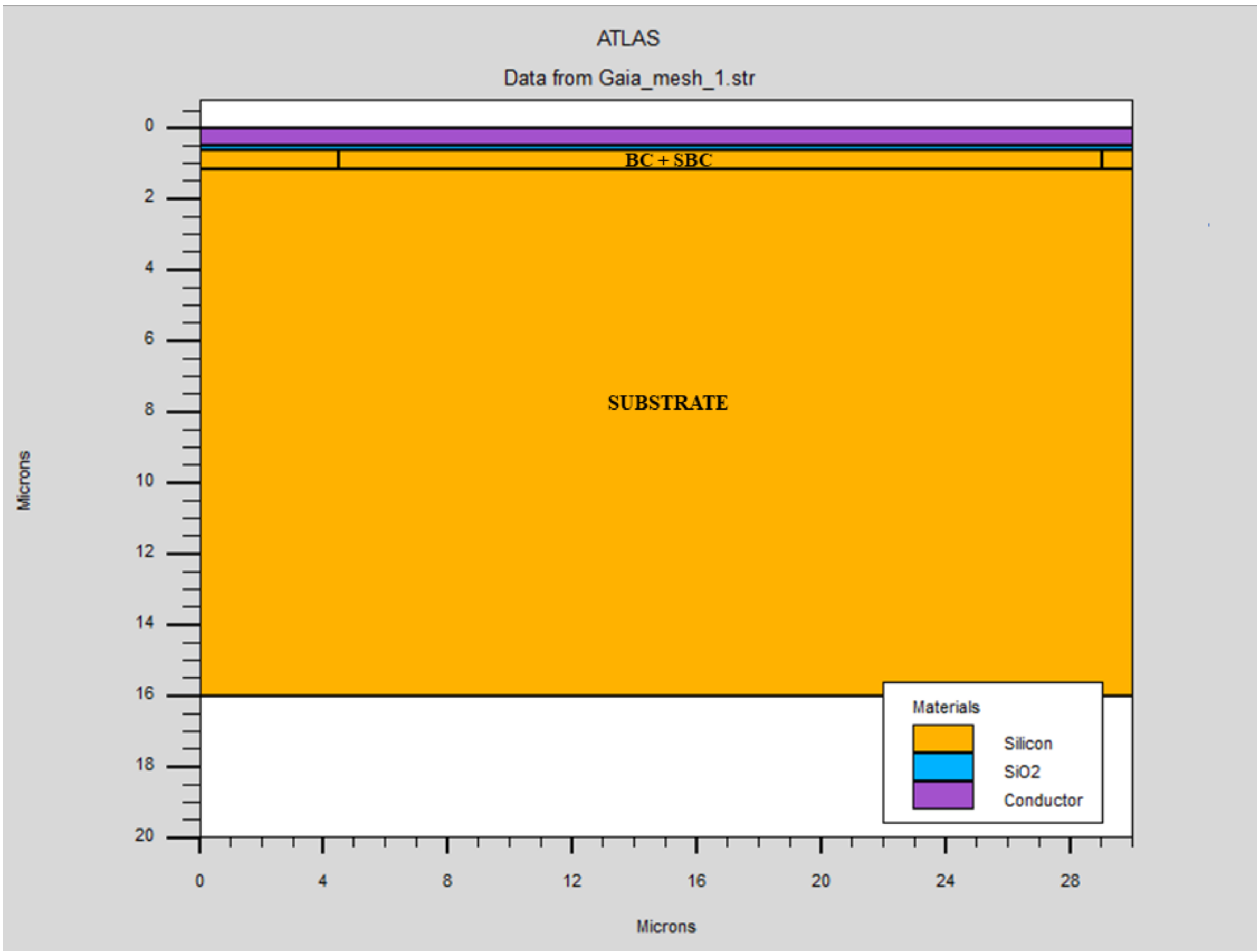

**Fig. 3.** The 2D Gaia AF CCD structure in the Across Scan (AC) direction as simulated in SILVACO ATLAS.

## 2.1 AR Coatings

We have elucidated the Gaia AF CCD pixel structure used in our simulations, but to test their efficiency the IQE can be used. To begin with, we have focused on single-layer AR coatings that have been applied on the backside of the CCD (substrate side). The thickness of the AR coating is estimated using the quarter wavelength formula (**Eqn. 1**).

$$AR\ coating\ thickness = \frac{\lambda}{4n} \qquad \textbf{(1)}$$

In **Eqn. 1**, λ is the wavelength at which the AR coating is centred, i.e., the wavelength at which the AR coating will allow for maximum photon absorption; n is the refractive index of the AR coating material at the wavelength λ (Lesser 1987). The SOPRA database of the SILVACO ATLAS software was tremendously helpful in allowing us to include the refractive indices of the AR coatings. When the refractive index of AR coatings at a specified wavelength comes in approximate agreement with $\sqrt{n}$, where "n" is the refractive index of the substrate material at that wavelength, we get excellent IQE values. To choose an AR coating this factor was deliberated.

On comparing the values of refractive indices of various AR coatings with the $\sqrt{n}$ , at different wavelengths, we got some AR coatings whose refractive indices match the $\sqrt{n}$ values and promise good results in IQE. In **Fig. 4** we can see the comparison of $HfO_2$ coating's refractive indices with the square root values of Si refractive indices. There is a very good matching which evidently explains why this AR coating is used in designing the Gaia AF CCDs.

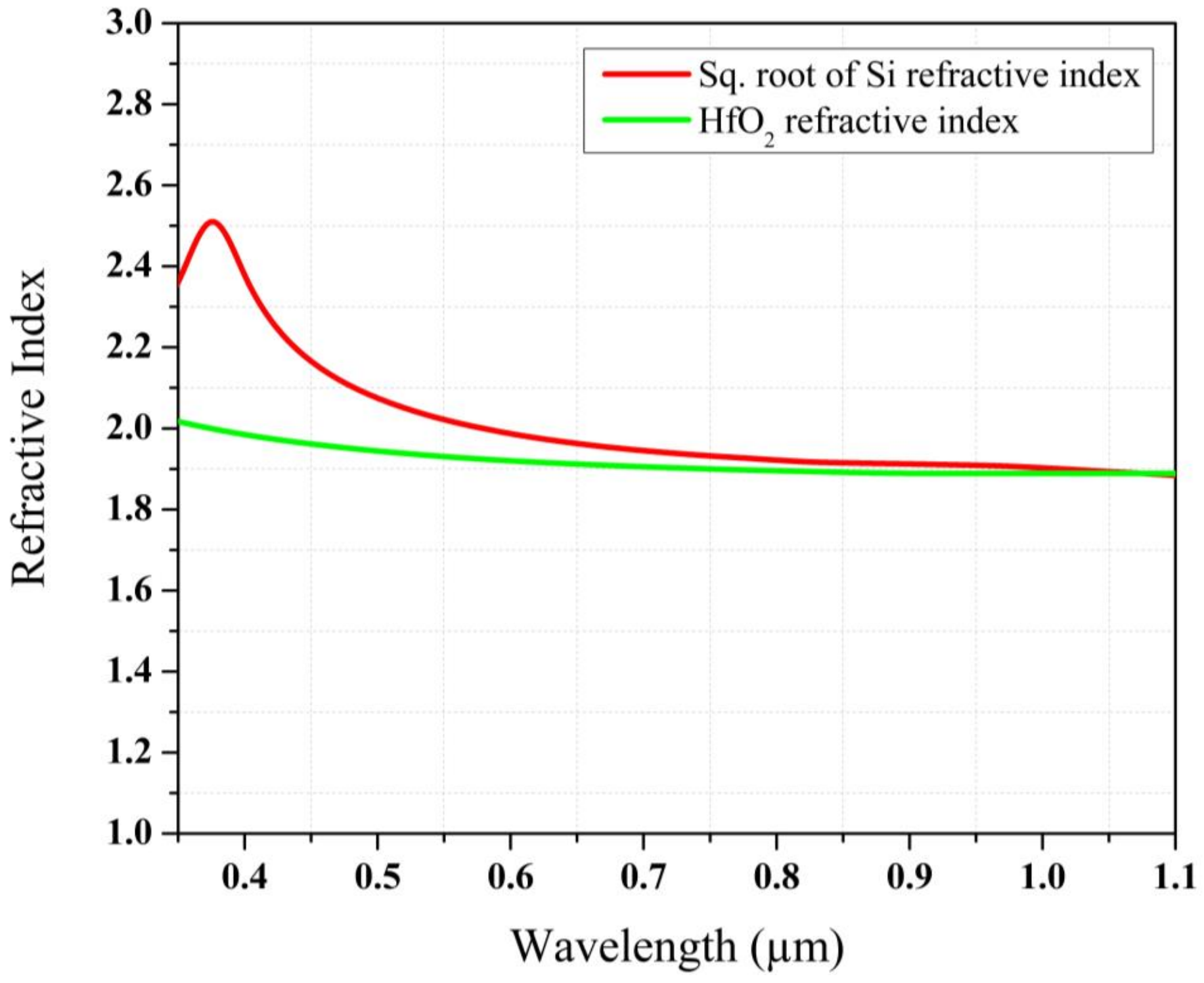

**Fig. 4:** Comparison of the refractive index of $HfO_2$ with the square root of the refractive index of Si at different wavelengths.

On performing the same comparison with various AR coatings, we got some promising results from ZnS, $Al_2O_3$, $ZrO_2$ and $Ta_2O_5$ which can be seen in **Fig. 5**. **Fig. 5(a)** shows how similar the values of $\sqrt{n}$ of Si and refractive index of ZnS are **Fig. 5(b)** shows that the refractive index of $Al_2O_3$ is also close to the $\sqrt{n}$ of Si values. **Fig. 5(c)** and **Fig. 5(d)** show similar results for $ZrO_2$ and $Ta_2O_5$.

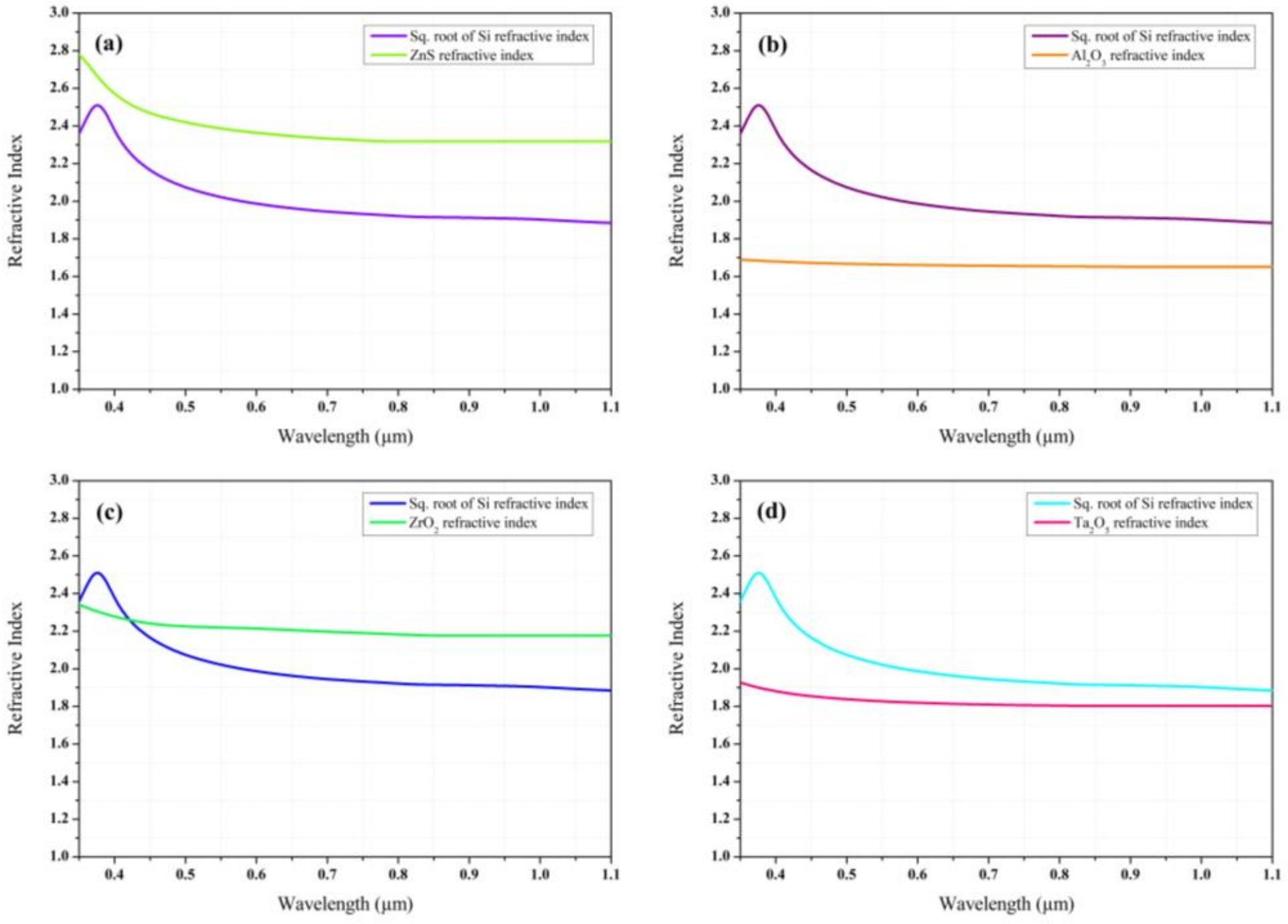

**Fig. 5:** Comparison of the square root of the refractive index of Si with **(a)** ZnS, **(b)** $Al_2O_3$, **(c)** $ZrO_2$, and **(d)** $Ta_2O_5$ at different wavelengths.

While choosing the materials for AR coatings their imaginary refractive index, which is a complex optical constant, often referred to as the extinction coefficient has also been considered. Using this we can calculate the absorption coefficient (α), which is the distance a photon can travel into the material before being absorbed. It is given by **Eqn. 2**.

$$\text{Absorption Coefficient } (\alpha) = \frac{4\pi k}{\lambda} \quad \textbf{(2)}$$

Where $\lambda$ is the wavelength of the incident light. We calculated the values of $\alpha$ for the whole wavelength range of the Gaia Astrometric Field (AF) CCD using SILVACO ATLAS with different AR coatings, which are presented in **Fig. 6 (a), (b), and (c)**. It can be concluded from these results, that the AR coatings have no effect on the absorption coefficient of silicon (Si).

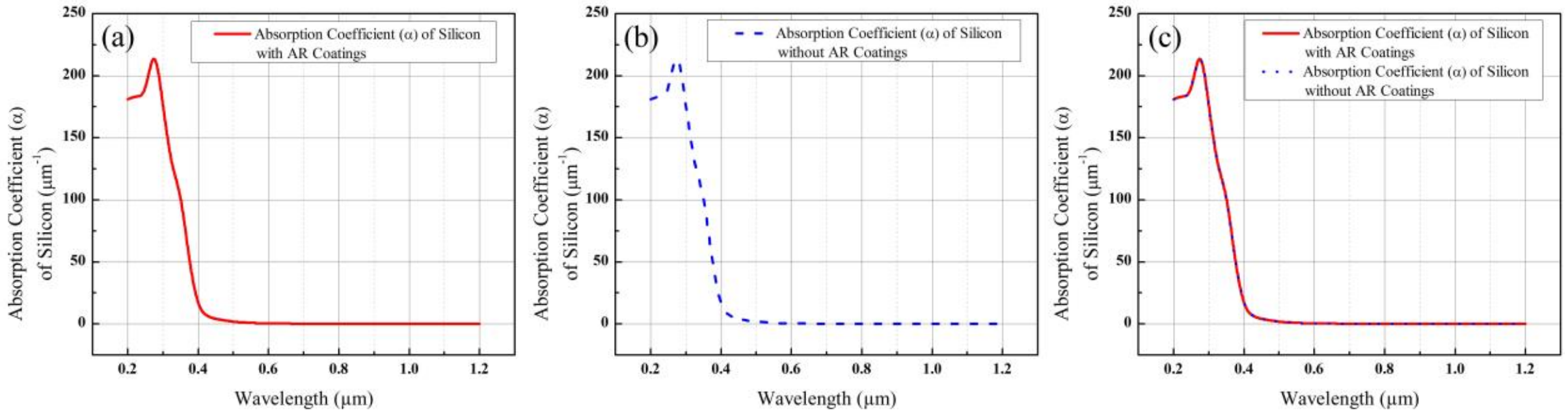

**Fig. 6:** The absorption coefficient of Si **(a)** with AR coatings, **(b)** without AR coatings and **(c)** with and without AR coatings.

Another factor that has been deliberated while selecting the AR coating material is reflectivity ($R_0$). **Eqn. 3** is used to calculate the reflectivity of a surface using the real refractive indices.

$$\text{Reflectivity } (R_0) = \left|\frac{n_1 - n_2}{n_1 + n_2}\right|^2 \quad \textbf{(3)}$$

Where $n_1$ and $n_2$ are the real refractive indices at a specific wavelength, of the medium from which the light is incident and the AR coating material respectively. The reflectivity of the Si-air interface is largely affected by the optical properties of the AR coating materials. The effect of AR coatings on the reflectivity of the Si-air interface is presented in **Fig. 7**, which have been calculated using the SILVACO ATLAS software.

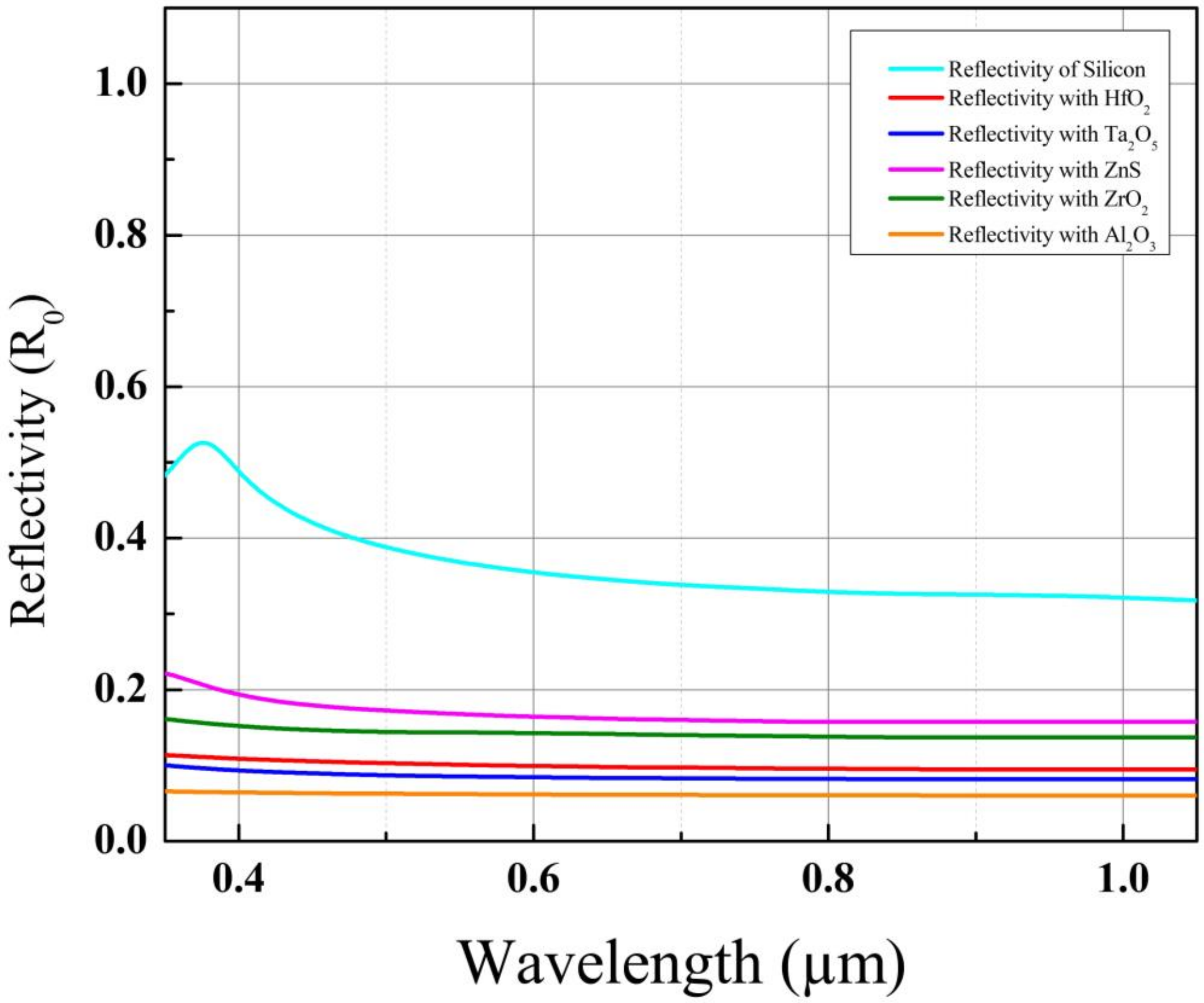

**Fig. 7:** Reflectivity of the CCD surface without any AR coating and with different AR coatings.

For an AR coating to perform well its respective refractive indices of the AR coating material also should be in approximate agreement with the square root of the respective refractive indices of the substrate material ($\sqrt{n}$) as per (Lesser 1987). $Ta_2O_5$ satisfies this condition the best among all the studied AR coating materials, and it also reduces the reflectivity substantially. Hence, it can be concluded that $Ta_2O_5$ can prove to be the best AR coating material for our case. Although $Al_2O_3$ reduces the reflectivity to the least value compared to all the studied AR coatings as presented in **Fig. 7**, its respective refractive indices are not very close to the square root of the respective refractive indices of Si (**Fig. 5 (b)**), which might hinder its performance.

## 3. Results and Discussion

While simulating the pixel structure with the $HfO_2$ AR coating, the **Eqn. 1** deduces the thickness of the coating as 0.085 µm. We identified the EQE for this setup to benchmark our simulations against the previously published EQE values by Walker et al. (2008). The simulation results are comparable to the experimental results, which can be visualised in **Fig. 8 (a)**. Any differences in the values of EQE might be because of the absence of ABD in our structure as the parameters required for its simulations were not published in the current literature.

On calculating the IQE for the Gaia AF CCD without applying any AR coating, it was discovered that they have a maximum IQE of 0.65 at the wavelength of 0.7 µm. Comparing the IQE of the CCDs with and without the $HfO_2$ coating, we clearly observe an average increment of about 37.2% in the whole wavelength range. **Fig. 8 (b)** coveys this increment graphically.

As per our study of AR coating materials, we choose four AR coatings to test our device. ZnS being one of them. Application of 0.069 µm thick single layer coating of ZnS, centred at 0.65 µm, unfolds that it is not very suitable for our device. The IQE curve in **Fig. 8 (c)** outlines that there is an average decrement of approximately 3.5% when studied parallel to the Gaia AF CCD with a $HfO_2$ AR coating in the whole range.

Our next choice was $Al_2O_3$ therefore we applied a single layer of it with a thickness of about 0.098 µm (calculated as per **Eqn. 1**), centred at 0.65 µm. The Gaia AF CCD with $HfO_2$ AR coating still provides better results for most wavelengths as compared to $Al_2O_3$. It is also observed that $Al_2O_3$ coating provides an average improvement in IQE of about 1.3% in the 0.35 µm to 0.425 µm range and an average decrement of about 1.9% in the rest of the wavelength range. These results are compiled in the graph presented in **Fig. 8 (d)**.

The next alternative that we analysed was a single-layer coating of $ZrO_2$ centred at 0.65

µm. The thickness of the coating was set as per **Eqn. 1** close to 0.073 µm. After observing the graphical data presented in **Fig. 8 (e)**, we infer that there is an average increment in the IQE of around 0.8% in the range of 0.4 µm to 0.525 µm when analogised with the IQE values when $HfO_2$ coating was applied. For the rest of the wavelengths, the AR coating of $HfO_2$ offers an average increment of almost 1.1%, when compared to $ZrO_2$.

Inspired by the improvements shown by $ZrO_2$, we adjusted its thickness to an approximate value of 0.071 µm, which allows for a peak absorbance of the light of wavelength 0.625 µm. This experimentation of ours yields an average increment of nearly 4% for the wavelength range of 0.375 µm to 0.575 µm when compared with $HfO_2$. with an average compromise of around 2% in the wavelength range of 0.575 µm to 1.05 µm. This data is visualised in **Fig. 8 (f)**.

We concluded our studies with $Ta_2O_5$ which is a promising AR coating. A single layer of $Ta_2O_5$, centred at 0.65 µm with a thickness of approximately 0.09 µm (calculated using **Eqn. 1**) permits almost the same performance as $HfO_2$. It is evident from **Fig. 8 (g)** that the two curves virtually overlap each other in the whole range. We did not obtain such promising results for any other AR coating. Hence, to improve the IQE in the low wavelength region, we deposited a 0.086 µm thick layer of $Ta_2O_5$ to centre the AR coating at 0.625 µm. The IQE of the Gaia AF CCDs with an AR coating of $Ta_2O_5$ shows an average increment of about 2.8% as compared to the CCDs with $HfO_2$ AR coating in the wavelength range of 0.375 µm to 0.6 µm. There is an average decrease of around 1.2% in the IQE of the wavelengths ranging from 0.6 µm to 1.05 µm which is clearly evident in **Fig. 8 (h)**.

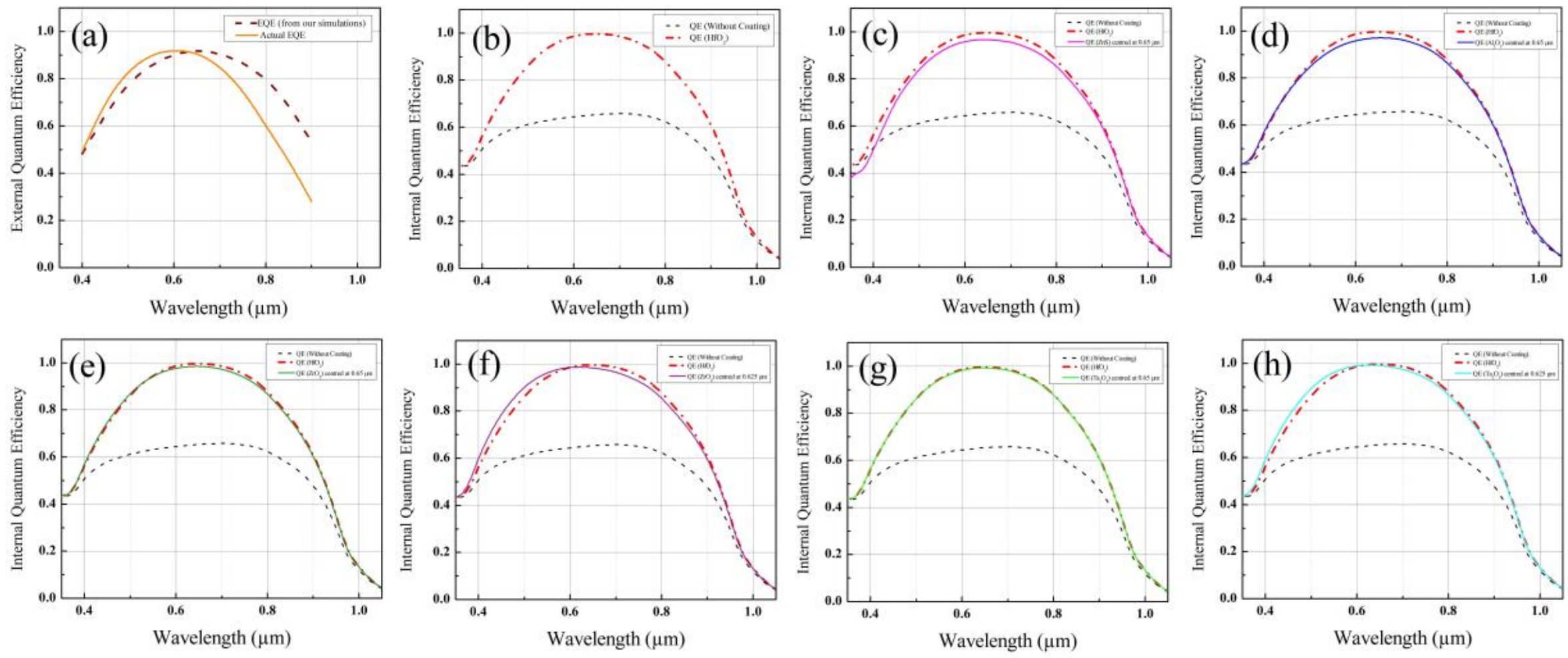

**Fig. 8.: (a)**The EQE of the Gaia AF CCDs derived from our simulations and the actual EQE reported by Walker et al. (2008). The IQE versus wavelength graph for Gaia AF CCD **(b)** without an AR coating and with a $HfO_2$ coating (centred at 0.65 µm), **(c)** ZnS coating (centred at 0.65 µm), **(d)** $Al_2O_3$ coating (centred at 0.65 µm), **(e)** $ZrO_2$ coating (centred at 0.65 µm), **(f)** $ZrO_2$ coating (centred at 0.625 µm), **(g)** $Ta_2O_5$ coating (centred at 0.65 µm), and **(h)** $Ta_2O_5$ coating (centred at 0.625 µm).

A comparative study has been drawn to analyse which AR coating best suits the Gaia AF CCDs. It is observed that a single layer coating of $ZrO_2$ and $Ta_2O_5$ gives similar or better results as compared to $HfO_2$ for most wavelengths lying in the Gaia's AF CCD range (330-1050 nm) (Seabroke et al. 2008). **Table 3** summarises the comparison among the IQE of CCDs for different AR coatings.

**Table 3:** Percentage increment and decrement in the QE of the Gaia AF CCD with different AR coatings as compared to the $HfO_2$ AR coating.

| S. N. | AR Coatings | Wavelength Range (µm) | Percentage Increment/Decrement in the IQE | Average Increment/Decrement in the IQE |
|---|---|---|---|---|
| **1.** | $HfO_2$ | 0.33 – 1.05 | Increment (2.6% to 53.3%) as compared to a CCD without any AR coating | Increment (37.2%) as compared to a CCD without any AR coating |
| **2.** | ZnS (centred at 0.65 µm) | 0.35 – 1.05 | Decrement (0.3% to 12.2%) as compared to $HfO_2$ | Decrement (3.5%) as compared to $HfO_2$ |
| **3.** | $Al_2O_3$ (centred at 0.65 µm) | 0.35 – 0.425 | Increment (0.1% to 2.3%) as compared to $HfO_2$ | Increment (1.3%) as compared to $HfO_2$ |
| | | 0.425-1.05 | Decrement (0.09% to 2.8%) as compared to $HfO_2$ | Decrement (1.9%) as compared to $HfO_2$ |
| **4.** | $ZrO_2$ (centred at 0.65 µm) | 0.4 – 0.525 | Increment (0.04% to 1.35%) as compared to $HfO_2$ | Increment (0.8%) as compared to $HfO_2$ |
| | | 0.525 -1.05 | Decrement (0.19% to 1.4%) as compared to $HfO_2$ | Decrement (1.1%) as compared to $HfO_2$ |
| **5.** | $ZrO_2$ (centred at 0.625 µm) | 0.375 – 0.575 | Increment (0.63% to 6.85%) as compared to $HfO_2$ | Increment (4%) as compared to $HfO_2$ |
| | | 0.575 – 1.05 | Decrement (0.38% to 2.72%) as compared to $HfO_2$ | Decrement (2%) as compared to $HfO_2$ |
| **6.** | $Ta_2O_5$ (centred at 0.65 µm) | 0.35 – 0.9 | Decrement (0.64% and less) as compared to $HfO_2$ | No prominent change as compared to $HfO_2$ |
| | | 0.9 - 1.05 | Increment (0.04% and less) as compared to $HfO_2$ | No prominent change as compared to $HfO_2$ |
| **7.** | $Ta_2O_5$ (centred at 0.625 µm) | 0.375 – 0.6 | Increment (0.17% to 5.15%) as compared to $HfO_2$ | Increment (2.8%) as compared to $HfO_2$ |
| | | 0.6 – 1.05 | Decrement (0.19% to 1.57%) as compared to $HfO_2$ | Decrement (1.2%) as compared to $HfO_2$ |

## Conclusions

We tested our Gaia AF CCD pixel model against the actual EQE values reported by Walker et al. (2008), which indicated the accuracy of our simulations. We presented our studies of the Gaia AF CCDs with various AR coatings to enhance the QE values. Any deviations in the results might be attributed to the absence of SBC and ABD in our structure since these features are proprietary to e2v. The performance of an AR coating can also be determined using a simple Si wafer. But, in order to validate our results for astronomical CCDs we used an electrical model in SILVACO ATLAS. This allows our results to be more acceptable to the astronomical community.

Our simulations establish that AR coatings are an important factor in improving the IQE and EQE of astronomical CCDs. They also elucidate that $Ta_2O_5$ and $ZrO_2$ are better materials for astronomical CCDs than $HfO_2$, mainly in the spectrum region from 0.330 µm to 0.575 µm. Hence, this work has important implications for the development of astronomical CCDs, which will help them to obtain the best possible data from satellites and telescopes. Multi-layer AR coatings can be deliberated for this purpose in the future which might enhance our knowledge of the Milky Way even more.

## Acknowledgements

1. The authors are grateful to Prof. Ashutosh Bharadwaj, University of Delhi, for his help and motivation throughout this research.
2. The authors are thankful to Dr. Harsupreet Kaur for providing us with the license of TCAD SILVACO ATLAS 2021 software.
3. The authors gratefully acknowledge Ms. Ritika Khatri, Delhi Technological University, for her continuous support and guidance during this research.